\documentclass[preprint2]{aastex}

\usepackage{subfigmat}
\usepackage{subfigure}
\usepackage{lineno}
\usepackage{color}

\newcommand{\ltsimscript}{\protect\raisebox{-0.5ex}{$\stackrel{\scriptstyle <}{\sim}$}} 
\newcommand{\rtsimscript}{\protect\raisebox{-0.5ex}{$\stackrel{\scriptstyle >}{\sim}$}} 

\slugcomment{Accepted for ApJ (30th Oct. 2016)}

\shorttitle{GRB141207A detected by Fermi}
\shortauthors{Arimoto et al.}

\begin{document}

\title{High-Energy Non-Thermal and Thermal Emission from GRB141207A detected by Fermi\\}


\author{Makoto Arimoto\altaffilmark{1,2}, Katsuaki Asano\altaffilmark{3}, Masanori Ohno\altaffilmark{4}, P\'eter Veres  \altaffilmark{5}, Magnus Axelsson  \altaffilmark{6,7}, Elisabetta Bissaldi\altaffilmark{8}, Yutaro Tachibana\altaffilmark{2} and Nobuyuki Kawai\altaffilmark{2}}
\email{m.arimoto@aoni.waseda.jp}


\altaffiltext{1}{Research Institute for Science and Engineering, Waseda University, 3-4-1, Ohkubo, Shinjuku, Tokyo, 169-8555, Japan}
\altaffiltext{2}{Tokyo Institute of Technology, 2-12-1 Ookayama, Meguro, Tokyo, 152-8551, Japan}
\altaffiltext{3}{Institute for Cosmic Ray Research, The University of Tokyo, 5-1-5 Kashiwanoha, Kashiwa, Chiba,  277-8582, Japan}
\altaffiltext{4}{Department of Physical Sciences, Hiroshima University, 1-3-1 Kagamiyama, Higashi-Hiroshima, Hiroshima, 739-8526, Japan}
\altaffiltext{5}{Center for Space Plasma and Aeronomic Research (CSPAR), University of Alabama in Huntsville, Huntsville, AL 35899, USA}
\altaffiltext{6}{KTH Royal Institute of Technology, Department of Physics, 106 91 Stockholm, Sweden}
\altaffiltext{7}{Tokyo Metropolitan University, Department of Physics, Minami-osawa 1-1, Hachioji, Tokyo 192-0397, Japan}
\altaffiltext{8}{Istituto Nazionale di Fisica Nucleare, Sezione di Bari, 70126 Bari, Italy}





\begin{abstract}
The bright long gamma-ray burst GRB 141207A was observed by the {\it Fermi Gamma-ray Space Telescope} and detected by both
instruments onboard. The observations show that the spectrum in the prompt phase is not well described by the canonical empirical Band function
alone, and that an additional power-law component is needed.
In the early phase of the prompt emission, 
 a modified blackbody with  a hard low-energy photon index ($\alpha$ = +0.2 -- +0.4) is detected, which suggests a photospheric origin. 
In  a finely time-resolved analysis, the spectra are also well fitted by the  modified blackbody combined with a power-law function. We  discuss the physical parameters of the photosphere such as  the bulk Lorentz factor of the relativistic flow and the radius. 
 We also discuss the physical origin of the extra power-law component observed during the prompt phase in the context of different models such as  leptonic and hadronic scenarios in the internal shock regime and  synchrotron emission in the external forward shock.
In the afterglow phase, the temporal and spectral behaviors of the temporally extended high-energy emission and the fading X-ray emission detected by XRT  on-board {\it Swift} are  consistent with  synchrotron emission in a radiative external forward shock. 
\end{abstract}





\section{Introduction}

Gamma-ray bursts (GRBs) are the brightest explosions in the universe.  During their early phases (i.e., prompt emission), most of their energy is released in the gamma-ray band and the typical duration is a few seconds to hundreds of seconds. The standard theoretical model for GRBs is the synchrotron shock model   \cite[see][for reviews]{2004RvMP...76.1143P};   gamma rays are emitted through non-thermal synchrotron emission via the internal dissipation of bulk kinetic energy of relativistic outflows, originating from a relativistically expanding fireball  \citep{1992MNRAS.258P..41R, 1994ApJ...430L..93R}. 

Most observational spectra are well fitted by an empirical function of two smoothly connected  power laws  \cite[e.g., Band function proposed by][]{1993ApJ...413..281B}. However, observations by the EGRET instrument onboard the {\it Compton Gamma-Ray Observatory} suggested 
 an additional spectral component in the high-energy band, which was not simply an extrapolation from the Band component during the prompt emission \citep{2003Natur.424..749G}. 
The {\it Fermi} observatory has achieved unprecedented broad-band sensitivity over seven decades in energy thanks to its two instruments: the Gamma-ray Burst Monitor (GBM) covering the energy band from 8 keV to 40 MeV \citep{2009ApJ...702..791M} and the Large Area Telescope (LAT) which is sensitive at the higher energy range from 20 MeV to  $>$300 GeV \citep{2009ApJ...697.1071A}. {\it Fermi} has  securely confirmed a distinct additional power-law component in several GRBs: e.g., GRB 090510 \citep{2010ApJ...716.1178A}, GRB 090902B \citep{2009ApJ...706L.138A}, GRB 090926 \citep{2011ApJ...729..114A}, GRB 110731B \citep{2013ApJ...763...71A}  and GRB 130427A \citep{2014Sci...343...42A}. 

 Although several models have been proposed to explain the mechanism behind the additional power-law component, its physical origin is still unclear. Several possible theoretical models exist:    synchrotron self-Compton in  internal shocks \citep[e.g.,][]{2010ApJ...720.1008C, 2011ApJ...739..103A},  hadronic cascade in internal shocks \citep[e.g.,][]{2010ApJ...725L.121A} and synchrotron emission from the external forward shock  \citep[e.g.,][]{2010MNRAS.409..226K}.
Furthermore, some GRBs have harder low-energy power-law indices  of the Band component than expected from the  synchrotron regime of the fast cooling case ($\alpha$ = $-$3/2) and  the slow cooling case ($\alpha$ = $-$2/3) \citep{1998ApJ...497L..17S}. This  is known as the synchrotron line of death \citep{1998ApJ...506L..23P}, and likely indicates that there exists a thermal component originating from the photosphere ($\alpha$ $\sim$ 1 if it is in the Rayleigh-Jeans regime).
From the theoretical point of view, ultra-relativistic expansion of the fireball predicts a thermal component in GRB spectra \citep{1986ApJ...308L..47G, 2002ARA&A..40..137M}. Indeed, a thermal component has been clearly identified in some GRBs by BATSE and {\it Fermi}.  While   thermal emission is sub-dominant compared with  non-thermal emission in most of the cases  \cite[e.g.,][]{2012ApJ...757L..31A, 2011ApJ...727L..33G, 2013ApJ...770...32G}, some GRBs have a dominant quasi-thermal component  from a few tens of keV to a few MeV \cite[e.g.,][]{2009ApJ...702.1211R, 2010ApJ...709L.172R}.

In this paper, we  present the observation and analysis of X-ray and gamma-ray emission from the long GRB 141207A by  the {\it Fermi} and {\it Swift}  instruments \citep{2004ApJ...611.1005G}  in  Sections \ref{Sec:Observation} and \ref{Sec:Analysis}.
The detailed characteristics of the obtained light curves and spectra  are described in  Sections \ref{Sec:Lightcurve} and \ref{Sec:SpectralAnalysis}, respectively.
The obtained spectrum of GRB 141207A has a statistically significant high-energy power-law component which cannot be extrapolated from the Band component, and  the dominant quasi-thermal emission is
detected in the early phase of the prompt emission.   The origin of the high-energy power-law component and the physical parameters of the thermal emission are discussed in Section \ref{Sec:Discussion}.  Furthermore, 
although its redshift is unclear, we estimated a possible redshift using an empirical correlation between the peak energy in the $\nu F_\nu$ space ($E_{\rm peak}$) and an isotropic equivalent luminosity $L_{\rm iso}$ \citep{2004ApJ...609..935Y, 2015ApJ...807..148G}. The obtained pseudo-redshift is very high ($z$ $\sim$ 10).

\section{Observations}\label{Sec:Observation}
On 2014 December  7, GRB 141207A triggered the {\it Fermi}-GBM instruments at 19:11:21.10 UT \citep{2014GCN..17150...1B}, which measured high-energy emission  up to 2 MeV. The burst's duration in the 50 -- 300 keV band was estimated to be $T_{\rm 90}$ = 21.0 $\pm$ 0.6 s  where $T_{\rm 90}$ is the time over which 5\% to 95\% of the total measured photons are detected. 
 The GBM location was found  by the on-ground calculation to be  R.A., Dec. = $161\fdg4$,  $+3\fdg2$ (J2000) with an error radius of $1\fdg1$ (1-$\sigma$ containment). It was at an angle of 59$^\circ$ from the LAT boresight at the time of the trigger. 
After the downlink and processing of the {\it Fermi}-LAT data, we found  significant emission above 100 MeV \citep{2014GCN..17146...1A},  which is  temporally correlated with the GBM emission. 
 Furthermore, the  LAT Low Energy (LLE) data from 20 MeV to  $\rtsimscript$ 100 MeV energies  also shows  emission coincident with that at higher energies in the LAT data at a 10-$\sigma$ level, where LLE data \citep{2010arXiv1002.2617P}  consist of events with a very loose event selection.  The calculated location using the LAT data was  found  to be R.A., Dec. = $159\fdg99$,   $+3\fdg91$ (J2000) with an error radius of  $0\fdg22$ (90\% containment), which is spatially consistent with the GBM position.
The LAT covered the beginning of the GBM emission and observed the burst until 1600 s after the trigger.
 GeV emission was significantly detected by the LAT, with 6 photons above 1 GeV in the initial 100 s and 
 a  highest-energy photon of 5.5 GeV observed 734 s after the GBM trigger.

 Target of Opportunity observations by the X-ray Telescope (XRT) and the UltraViolet and Optical Telescope (UVOT) onboard {\it Swift}  began $\sim$13 hours after the GBM trigger. A fading X-ray source was detected by 
XRT \citep{2014GCN..17157...1E} and the obtained location was R.A., Dec. = $159\fdg8547$,  $+3\fdg7114$ (J2000) with an error radius of  $0\fdg0014$ (90\% containment), which is consistent with the LAT position.
While the X-ray counterpart was detected, no significant emission in the {\it u} band was confirmed by UVOT. 
 Although a ground-based
 optical observation was performed 16 hours after the trigger  \citep{2014GCN..17152...1T}, no optical source was detected because the fading X-ray source was outside the FoV, and a redshift could therefore not be measured.
 
\section{Data analysis}\label{Sec:Analysis}
For the temporal and spectral analyses of the burst, we used GBM and LAT data obtained from the {\it Fermi} Science Support Center server and followed the standard analysis procedure \footnote{http://fermi.gsfc.nasa.gov/ssc/data/access/} using {\it Fermi} Science tools.
For the GBM analysis, we used three NaI detectors within a burst angle of 60$^\circ$ (NaIs 0, 3 and 4) and the brightest BGO detector (BGO 0). 

For the LAT data, Instrument Response Functions (IRFs) were recently greatly updated by improving knowledge of the LAT detector and the operation environment 
\citep[Pass8 \footnote{http://fermi.gsfc.nasa.gov/ssc/data/analysis/documentation/Pass8\_usage.html};][]{2013arXiv1303.3514A}.
As a result, a significant reduction of background events and increase of effective area were accomplished.
In this paper, we used data with ``{\tt P8TRANSIENT010E"} and ``{\tt P8SOURCE"} instrument response functions (IRFs) when analyzing the prompt emission and extended emission (e.g., early afterglow), respectively. 
The {\tt P8TRANSIENT010E} data come from a loose-cut filter optimized for short-time transients ($\sim$100 s) under background-suppressed conditions and {\tt P8SOURCE} data has a tighter filter than {\tt P8TRANSIENT010E}, which is suitable for analysis of longer time intervals ($\geq$ 1000 s) to reduce the contribution of background events (e.g., charged particles, cosmic rays etc.).

We also performed an analysis using {\it Swift} XRT data. To extract the light curve and spectrum, we utilized the method of automatic analysis based on \cite{2009MNRAS.397.1177E}.

\section{Light curves}\label{Sec:Lightcurve}
\begin{figure}[t]
\epsscale{1.0}
\plotone{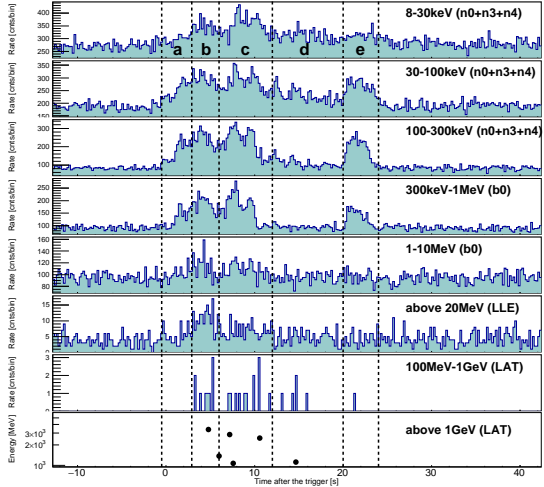}
\caption{Composite light curve of GRB141207A. The top five panels are different bands in the GBM, and the lower three show LAT photons. \label{compositeLC}}
\end{figure}

We show the composite light curves including the GBM and LAT data in Figure \ref{compositeLC}.
 The observed light curves clearly show that the gamma-ray emission  above  20 MeV detected by the LAT is delayed by $\sim$3.4 s with respect to  the GBM emission at energies $\ltsimscript$ 1 MeV. 
 This delay is also  suggested by the light curve of the 1 -- 10 MeV band of the BGO detector and is a common feature observed in other GRBs 
 \citep{2009Sci...323.1688A, 2009ApJ...706L.138A, 2009Natur.462..331A, 2009ApJ...707..580A, 2010ApJ...712..558A, 2010ApJ...716.1178A, 2011ApJ...729..114A, 2013ApJ...763...71A}. 
 The  highest-energy photons in the initial 100\,s interval and the whole interval until $T_0$ + 1600 s, where the significant GeV emission was detected, are 3.4 GeV and 5.5 GeV observed at $T_0$ + 4.8 s  and $T_0$ + 734 s, respectively.  To estimate how well these photons are associated with the GRB,  we calculated chance probabilities that these two photons originate from background  events such as Galactic or extragalactic diffuse emission to be 1.16$\times$10$^{-3}$ and 1.47$\times$10$^{-4}$ respectively,  using the {\tt gtsrcprob} analysis provided by  the {\it Fermi} Science tools. Thus, the detected highest-energy photons are very likely to originate from GRB141207A.
The standard $T_{90}$ duration  in the GBM energy range (50 -- 300 keV)  is 21.0 $\pm$ 0.6 s and we define the time interval where the GBM emission clearly appeared (from $T_0$ $-$ 0.5 s to + 24 s) as a prompt emission phase.
 In the prompt emission phase, the duration in the LAT band is almost the same as or slightly shorter than in the GBM band, although the number of the LAT photons is small and there is a large uncertainty due to photon statistics.

 In order to perform time-resolved spectroscopy to characterize the spectral evolution of this burst, we chose  five time intervals,  which are indicated by vertical lines in Figure \ref{compositeLC}. 

In the first time interval (from $T_0$ $-$0.5 s to +2.9 s, labeled ``$a$"), while there were no LAT photons, significant emission is present in the GBM. The second time interval  (from $T_0$ + 2.9 s to 6 s, ``$b$") contains the first bright LAT  emission episode coincident with the GBM pulse. In the third time interval (from $T_0$ + 6 s to 12 s, ``$c$") there were prominent GBM pulses and LAT emission. In the fourth time interval  (from $T_0$ + 12 s to 20 s, ``$d$") there was no emission in the high-energy GBM band ($\geq$300 keV), yet emission in the low-energy GBM band (8 -- 100 keV) and LAT emission were detected. In the fifth time interval  (from $T_0$ + 20 s to 24 s, ``$e$")  a rebrightening of the GBM emission is seen while there are few LAT photons.

\begin{figure}[t]
\epsscale{1.0}
\plotone{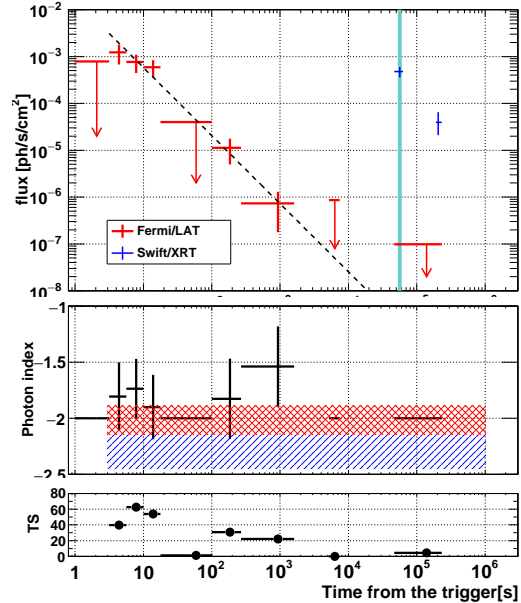}
\caption{Flux and photon index for the LAT extended emission and afterglow flux detected by the XRT for GRB 141207A (LAT:  100 MeV -- 10 GeV, XRT: 0.3 -- 10 keV). The statistical significance is given by $\sim$$\sqrt{TS}$. The photon index is fixed to $-$2 when the upper limit of the flux is calculated due to low significance ({\it TS} $<$ 10). The dashed line represents the best-fit power-law function in the LAT data.  The  thick vertical   cyan  line defines the time interval of the joint LAT-XRT SED discussed in Section \ref{sec:afterglow}.  Hatched blue and cross-hatched red pattern areas show the 1-$\sigma$ permitted regions in adiabatic- and radiative-jet cases, respectively, as discussed in Section \ref{sec:afterglow}. \label{EE_LAT_Swift}}
\end{figure}

Figure \ref{EE_LAT_Swift} shows the extended emission detected by LAT at longer timescales than the prompt phase,  until 1600 s after the GBM trigger time. When calculating the 
LAT flux, we used the test statistic  \citep[$TS$;][]{1996ApJ...461..396M} to estimate the statistical significance and test whether LAT emission was significant or not. The significance in units of the standard 
deviation scales roughly as $\sqrt{TS}$. The long-lasting emission was detected until 1600 s after the trigger with a confidence level greater than 4.5 $\sigma$.
The LAT flux decreases with time and follows a power-law function of  $t^{-1.5\pm0.2}$. The temporal index is similar to other GRBs \citep{2013ApJS..209...11A}.
The photon index is almost constant with values between $-$1.5 and $-$2.0. 
XRT also detected the fading X-ray emission and its temporal decay index is $-$1.9$\pm$0.6, which is  consistent with the LAT temporal index within 1-$\sigma$ uncertainty.

\section{Spectral analysis}\label{Sec:SpectralAnalysis}
 We performed time-integrated and time-resolved spectral analyses with the GBM and LAT data for the defined time intervals and show comparisons of fit results for each time interval 
 in Table \ref{spec_result_exp_comp}.
 When performing spectral analysis, we used {\it XSPEC} version 12.8.2  \citep{1996ASPC..101...17A}.
 For the fitting procedure, we adopted the ``PGstat" statistic, suitable for low count statistics \citep{2011hxra.book.....A}.
 
 \subsection{Time-integrated spectrum}\label{Sec:TimeResolvSpec}
\begin{figure}[h]
\epsscale{1.00}
\plotone{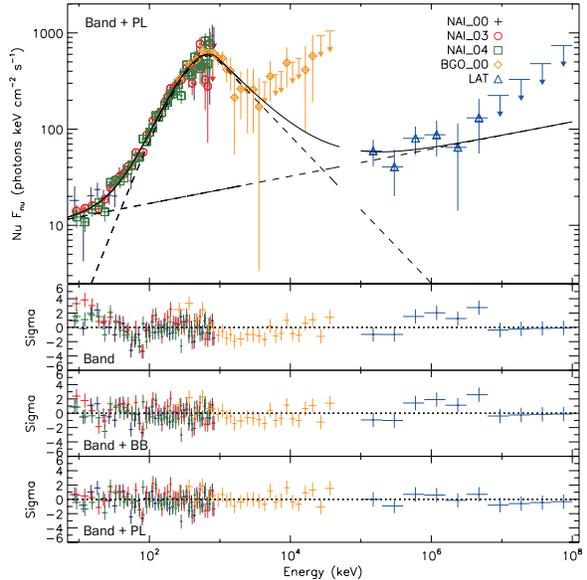}
\caption{Time-integrated spectra (from $T_0$ $-$0.5 s -- +24.0 s) observed with GBM and LAT. The upper panel shows the best-fitting model (the Band plus power-law functions). Residuals of three models are shown in the lower three panels; $Top$: the Band function, $Middle$: the Band function plus blackbody (BB), $Bottom$: the Band plus power-law (PL) functions. \label{specSummaryBand}}
\end{figure}

 First, we performed the time-integrated spectral analysis during $T_0$ $-$ 0.5 s to +24 s, denoted as ``total" in Table \ref{spec_result_exp_comp}.  We fitted these initial  24.5 s of the prompt emission using the empirical Band function and obtained a fluence of (7.47 $\pm$ 0.30) $\times$ 10$^{-5}$  erg cm$^{-2}$ over the whole GBM + LAT energy range (8 keV -- 10 GeV).   The obtained spectrum is shown in Figure \ref{specSummaryBand}. Three models - the Band function, the Band function plus blackbody (BB) and the Band plus power-law (PL) functions - were tested.
 Fitting only the Band function shows systematic bumps around 10 keV and a few GeV as shown in the residuals of the Band  function-only fits of Figure \ref{specSummaryBand}. This indicates that an additional spectral component  is needed. Thus, to mitigate the residuals of the low-energy wavy structure, a BB component was added. The statistical improvement (i.e., goodness of fit) is qualitatively estimated by the difference in PGstat values, hereafter denoted as $\Delta$PGstat, and by adding the BB component a drastic improvement of $\Delta$PGstat = 72 for  an increase of 2 degrees of freedom (Band vs. Band+BB) is obtained, although the residuals around a few GeV still remain as a bump as shown in the residual of  Band + BB  in Figure \ref{specSummaryBand}.
The remaining high-energy bump indicates that there could exist an extra high-energy component. We next adopted the PL function as an extra component instead of the BB component to reduce the residual of both the low-energy and high-energy bumps as shown in the residual of  Band + PL  in Figure \ref{specSummaryBand}.
The obtained statistical improvement of $\Delta$PGstat = 97 (Band vs. Band+PL) is slightly better than the case of Band vs. Band+BB. 
Considering both cases, the obtained spectrum cannot be well explained by only a Band function and an extra component is clearly needed.
However, the improvement of $\Delta$PGstat = 25 (Band+BB vs. Band+PL) is not large compared with the previous cases (i.e., Band vs. Band+BB or Band vs. Band+PL) and we need a null hypothesis probability to estimate how significant Band+PL is relative to Band+BB. 
 
 To estimate the significance of the power-law component we use a Monte-Carlo simulation. The explicit procedure is the following: (1) First, we derived a best-fit function as a null hypothesis (Band+BB) from the obtained spectrum. (2) Using the derived best-fit parameters, dummy spectra were created with the {\it fakeit} command of {\it XSPEC}. In this paper, we tried 1 million realizations. (3) We obtained the best-fit parameters with the null hypothesis model (i.e., Band+BB) for each realization. (4) The same fitting with an alternative model (e.g., Band+PL) was performed. (5) We calculated $\Delta$PGstat between the null hypothesis and alternative models for each realization. Finally, this gives the reference probability distribution.    $\Delta$PGstat = 25 corresponds to a probability of $\sim$10$^{-5}$ that the best-fit function will be Band+PL if the underlying function is Band+BB.  We thus can say that Band+PL is a more likely model to represent the obtained time-integrated spectrum.
 
 Although we tried a three-component model (i.e., Band+PL+BB), we did not obtain any statistically significant improvements ($\Delta$PGstat $\le$ 3) and in particular the additional BB parameters are not well constrained. Furthermore, we tried the three-component function at several separated time intervals as mentioned in Sec. \ref{SubSec:CoarseTimeResolvSpec} and \ref{Sec:FineTimeResolvedSpec}. However,  we confirm that there is no  significant improvement in any time interval by adopting the three-component model.  
 
  \begin{figure*}[t]
\epsscale{2.0}
\plotone{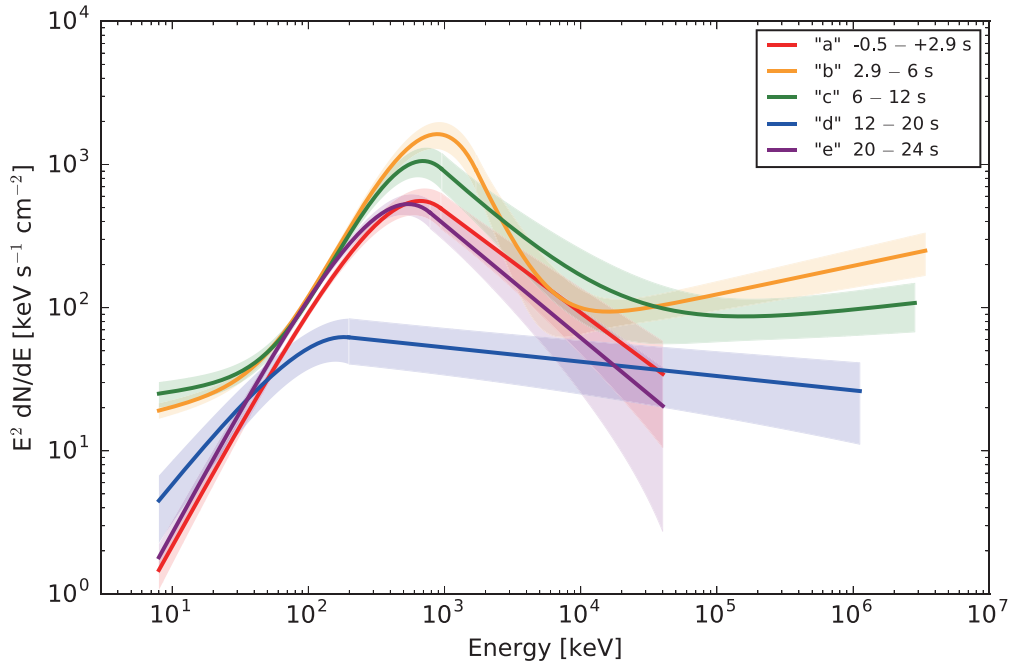}
\caption{Coarse time-resolved spectra in the defined time intervals. In the very early interval ``$a$" ($T_0$ $-$ 0.5 s to +2.9 s) the Band function can model the obtained spectrum well, after which the  significant extra power-law component emerges (Band + PL) in the time intervals ``$b$" (from $T_0$ + 2.9 s to 6.0 s) and ``$c$" (from $T_0$ + 6.0 s to 12.0 s) and a soft-to-hard evolution is seen. In  the time intervals ``$d$" (from $T_0$ + 12.0 s to 20.0 s) and ``$e$" (from $T_0$ + 20.0 s to 24.0 s) there  is no significant extra component. The Band function is enough to explain the obtained spectrum and the spectrum softens  (hard-to-soft evolution).   The shaded areas correspond to the 1-$\sigma$ confidence region. \label{spec_time_resolved}}
\end{figure*}
 
The low-energy photon index of the Band component is large ($\alpha$ $\sim$ $-$0.1) and such a hard index cannot be explained by the synchrotron emission model which limits the expected photon indices to $\alpha$ $<$ $-$3/2 or $-$2/3 for the  fast- and slow-cooling cases, respectively  \citep{1998ApJ...497L..17S}.
The obtained photon index suggested that the Band component should not originate from the synchrotron emission by accelerated electrons but  instead from a thermal process; the low-energy photon index would  then relate  to the spectral slope in  the Rayleigh-Jeans regime of the blackbody component ($\alpha$ = +1). 
 
\begin{table*}[h]
\begin{center}
\caption{Spectral fitting with LAT and GBM in the defined time intervals.\label{spec_result_exp_comp}}
\footnotesize 
\begin{tabular}{clcccccc}
\tableline\tableline
Interval & Model & $\alpha$, $p$ & $\beta$ & $E_{\rm peak}$ & $kT_{\rm ob}$ & $\Gamma_{\rm ext}$ & PGstat/dof 
\\
{[s]}  & & &  & [keV] & [keV] &  &   \\
\tableline
 & Band & $-$0.66$\pm$0.05 & $-$2.45$^{+0.05}_{-0.06}$  & 940$^{+102}_{-87}$ & --- &  --- & 530/482   \\
``total" & Band+BB & $-$0.33$^{+0.13}_{-0.12}$ & $-$2.41$\pm0.05$  & 702$^{+82}_{-67}$ & 5.6$^{+1.2}_{-1.3}$ &  --- & 458/480 \\
$-$0.5 -- +24.0  & Band+PL & $-$0.08$^{+0.27}_{-0.17}$  &$-$2.87$^{+0.35}_{-1.01}$  & 688$^{+69}_{-61}$ & ---  & $-$1.85$^{+0.03}_{-0.04}$  & 433/480  \\   \hline
 & Band & $-$0.28$^{+0.20}_{-0.14}$ & $-$2.94$^{+0.62}_{-unc}$  & 666$^{+125}_{-99}$ & --- &  --- & 464/476   \\
``$a$" & CUTPL & $-$0.23$^{+0.15}_{-0.20}$ & ---  & 661$^{+143}_{-89}$ & --- &  --- & 466/477 \\
$-$0.5 -- +2.9  & CUTPL+BB & $-$0.15$^{+0.23}_{-0.56}$  & ---  & 637$^{unc}_{-93}$ & 8.6$^{+unc}_{-unc}$  & ---  & 465/475  \\
 & CUTPL+PL & $-$0.19$^{+0.40}_{-0.16}$  & ---  & 643$^{+172}_{-104}$ & ---  & $-$1.41$^{+unc}_{-unc}$  & 465/475  \\ \hline
 & Band & $-$0.46$^{+0.05}_{-0.08}$ & $-$2.50$^{+0.09}_{-0.05}$  & 1128$^{+143}_{-121}$ & --- &  --- & 562/482   \\
``$b$" & Band+BB & +0.02$^{+0.27}_{-0.19}$ & $-$2.46$^{+0.09}_{-0.10}$  & 848$^{+110}_{-104}$ & 7.5$^{+2.2}_{-2.1}$ &  --- & 504/480 \\
2.9 -- 6.0  & Band+PL & +0.24$\pm0.21$  &$-$4.03$^{+1.03}_{-unc}$  & 878$^{+106}_{-81}$ & ---  & $-$1.81$^{+0.04}_{-0.05}$  & 476/480  \\   
 &  DISKPBB+PL &  1.10$^{+0.27}_{-0.14}$  & --- & --- &  328$^{+44}_{-38}$ &  $-$1.80$^{+0.05}_{-0.05}$  &  477/481\\ \hline

  & Band & $-$0.67$^{+0.06}_{-0.07}$ & $-$2.50$^{+0.08}_{-0.10}$  & 1169$^{+170}_{-128}$ & --- &  --- & 605/482   \\
``$c$" & Band+BB & $-$0.04$^{+0.19}_{-0.16}$ & $-$2.43$^{+0.07}_{-0.08}$  & 711$^{+90}_{-78}$ & 5.7$^{+0.8}_{-0.7}$ &  --- & 461/480 \\
6.0 -- 12.0  & Band+PL & +0.37$^{+0.29}_{-0.20}$  &$-$2.88$^{+0.34}_{-0.82}$  & 692$^{+80}_{-81}$ & ---  & $-$1.89$^{+0.07}_{-0.04}$  & 448/480  \\   
 &  DISKPBB+PL &  1.18$^{+0.35}_{-0.17}$  & --- & --- & 268$^{+35}_{-30}$ &  $-$1.87$^{+0.04}_{-0.05}$  &  456/481  \\ \hline

   & Band & $-$0.79$^{+0.31}_{-0.23}$ & $-$2.1$^{+0.07}_{-0.10}$  & 182$^{+68}_{-43}$ & --- &  --- & 539/482   \\
``$d$" & Band+BB & $-$0.48$^{+0.72}_{-0.38}$ & $-$2.1$^{+0.08}_{-0.10}$  & 155$^{+57}_{-37}$ & 2.1$^{+1.7}_{-1.5}$ &  --- & 534/480 \\
12.0 -- 20.0  & CUTPL+PL & +0.31$^{+0.81}_{-0.84}$  & --- & 169$^{+41}_{-26}$ & ---  & $-$1.87$^{+0.04}_{-0.07}$  & 530/480  \\   \hline
  & Band & $-$0.25$^{+0.18}_{-0.13}$ & $-$2.79$^{+0.27}_{-0.48}$  & 551$^{+70}_{-77}$ & --- &  --- & 484/482   \\
``$e$" & Band+BB & $-$0.11$^{+0.41}_{-0.19}$ & $-$2.77$^{+0.26}_{-0.45}$  & 505$^{+78}_{-68}$ & 3.7$^{+7.4}_{-2.0}$ &  --- & 479/480 \\
20.0 -- 24.0  & Band+PL & +0.01$^{+0.27}_{-0.23}$  &$-$9.37$^{fixed}$  & 518$^{+64}_{-60}$ & ---  & $-$1.96$^{+0.16}_{-0.22}$  & 477/480  \\   \hline

\tableline
\end{tabular}
\tablecomments{The Band function parameters are the low-energy photon index $\alpha$, the high-energy photon index $\beta$ and the peak energy $E_{\rm peak}$. The blackbody (BB) 
 or phenomenological modified blackbody (DISKPBB) 
temperatures are indicated as $kT_{\rm ob}$. $p$  denotes the exponent of the radial dependence of the disk temperature while $\Gamma_{\rm ext}$ denotes the photon index of the extra power-law component (PL). For DISKPBB, the photon index of the Rayleigh-Jeans part corresponds to 2 - 2/$p$. We  use a power-law function with exponential cutoff (CUTPL) instead of the Band function when the high-energy photon index of the Band component is very steep ($<$ $-$3) and unconstrained. Errors correspond to the 90\% confidence region. ``$unc$" means that the fitting parameter is unconstrained. }
\end{center}
\end{table*}
 
 \subsection{Coarse time-resolved spectral analysis}\label{SubSec:CoarseTimeResolvSpec}
In this section, we describe the spectral evolution with time for the defined time intervals. Values from all fits are
shown in Table~\ref{spec_result_exp_comp}. As described below, strong spectral evolution is seen in the prompt phase.

 In the first time interval {\it a},  the Band function or a power-law function with exponential cutoff (CUTPL) provides a good fit to the data and there is no statistical difference between Band and CUTPL. 
This is also consistent with the fact that in the case of fitting the Band function  alone  the high-energy photon index $\beta$ is very low ($\beta$ $<$ $-$3). 
 Furthermore, when we adopt CUTPL to represent the obtained spectrum there is no need for any additional spectral component, 
 as there is no statistical improvement by adding an extra component such as the blackbody or a power-law function ($\Delta$PGstat $\sim$ 1) as shown in Table \ref{spec_result_exp_comp}.

In the second and third time intervals, {\it b} and {\it c},   if we choose only the Band function, 
 we find the same residuals as obtained in Sec. \ref{Sec:TimeResolvSpec} (e.g., the low-energy bump). We see that the best-fit model is Band+PL and there is clear statistical improvement with respect to the case of the Band function  alone ($\Delta$PGstat = 86 and 157 in intervals {\it b} and {\it c}, respectively). Compared with the spectrum in the initial interval, the Band component low-energy index becomes harder. Interestingly, in  these intervals the obtained low-energy photon indices are too large ($\alpha$ = +0.2 -- +0.4) to be explained by non-thermal synchrotron emission as described in Section \ref{Sec:TimeResolvSpec}. Together with the small high-energy photon indices ($\beta$ $\leq$$-$3), this indicates a narrow spectral component originating in thermal emission from the photosphere.

From the theoretical point of view, \cite{2010MNRAS.407.1033B} 
pointed out that synchrotron emission and temporal variability lead to a
softer spectrum ($\alpha$ = +0.4) than the Rayleigh-Jeans index ($\alpha$ = +1).
In addition, a relativistically expanding conical jet
structure implies a dispersion in the Doppler factor, which can broaden the photon
spectrum \citep{2013JCAP...09..008A}.
 Here, we choose the {\tt diskpbb} model implemented in {\it XSPEC} as a modified blackbody model to reproduce the thermal photospheric emission. The {\tt diskpbb} model represents  a multi-temperature blackbody spectrum where the local disk temperature $T(r)$ at a distance $r$ from a compact object  is proportional to $r^{-p}$. While this 
model describes optically thick radiation from a geometrically thin accretion disk, we introduce this model {\it phenomenologically} to reproduce our modified blackbody spectrum. If we adopt this model, the photon index in the Rayleigh-Jeans range is represented as 2 - 2/$p$ and the {\tt diskpbb} model can obtain a broader spectrum than a  pure blackbody. The {\tt diskbb} model plus the power-law function, denoted as DISKPBB+PL,  gives a fit as good as the case of Band+PL  in intervals {\it b} and {\it c}.

In the fourth and fifth time intervals, {\it d} and {\it e}, as  $\Delta$PGstat of Band vs. Band+BB or Band vs. Band+PL is small ($\Delta$PGstat $<$10) the additional power-law component is not needed statistically. If we adopt only a Band function to describe the spectrum, it has a smaller low-energy index and becomes softer, matching the hard-to-soft evolution 
common for typical GRBs.
However, in the fourth time interval ({\it d}),  the value of $\beta$ ($\sim$$-$2.1) obtained from the simple Band function fit is larger than that of typical GRBs \citep{2006ApJS..166..298K}, which might  imply that there exists a hidden high-energy power-law component, although the statistical significance is low.
   
We show the best-fit energy spectra in the five  time intervals in Figure \ref{spec_time_resolved}. We note that the additional power-law components are displayed only in intervals {\it b} and {\it c} while in the other intervals there is no significant improvement to claim the additional component.

\begin{figure*}[h]
\centering
\epsscale{2.00}
\plotone{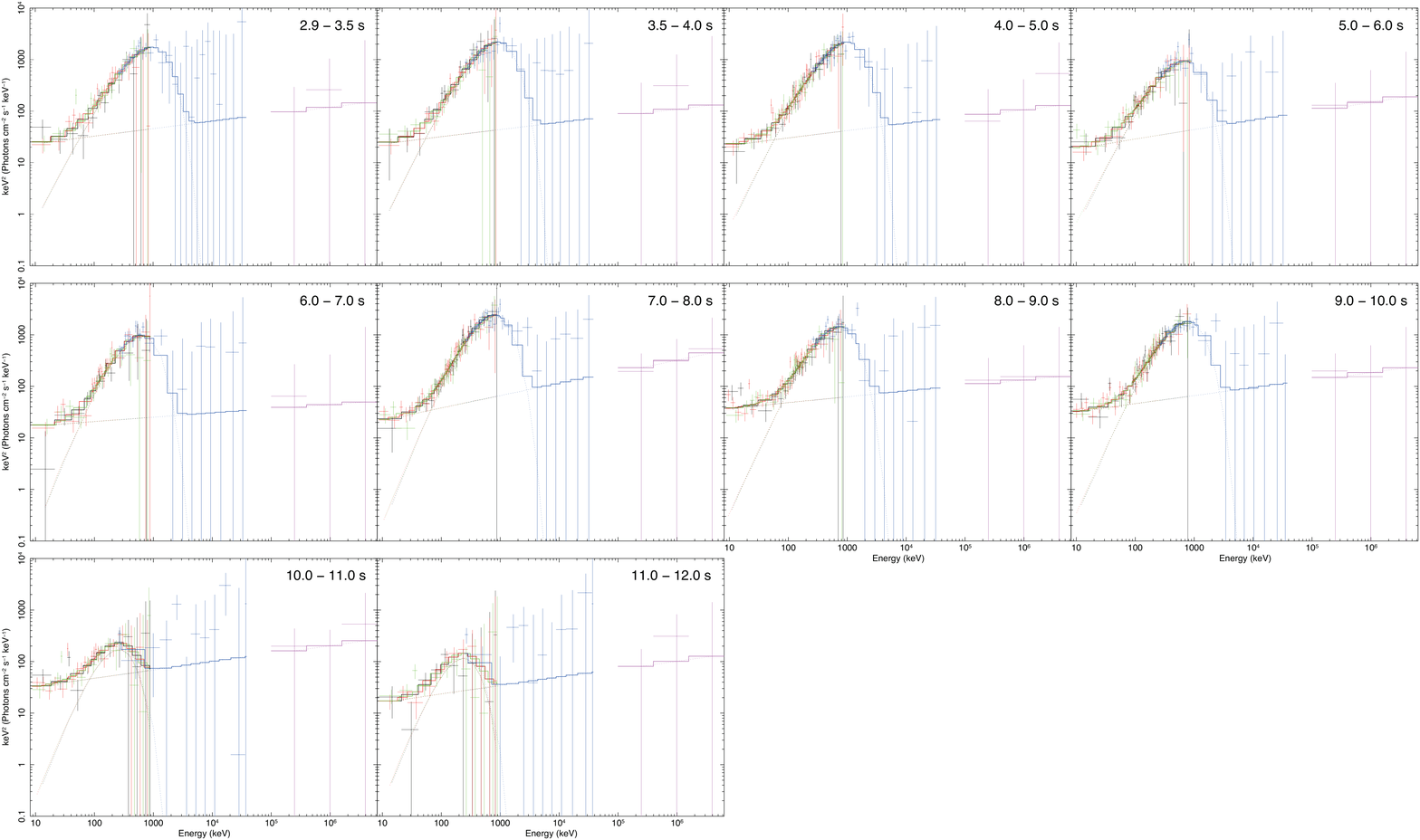}
\caption{Fine time-resolved spectra fitted with   the modified BB (diskpbb) + PL function in the time interval from 2.9 s to 12.0 s after the trigger (intervals ``$b$" and ``$c$" where the additional power-law component is significantly detected).
}\label{fig:spec_time_resolved_pow_BB}
\end{figure*}

\subsection{Fine time-resolved spectral analysis in time intervals {\it b} and {\it c}}\label{Sec:FineTimeResolvedSpec}

To characterize the photospheric emission,  we used  the modified blackbody model 
plus a power-law function, as previously shown in Sec. \ref{SubSec:CoarseTimeResolvSpec}.
 The  previously selected time intervals {\it {\rm (}c {\rm and} d{\rm )}} during which the spectra are well represented by Band+PL  or DISKPBB+PL,  contain several pulses with  variability timescales of a few seconds and  the previous divisions are too coarse to resolve the temporal variation.
 To derive a minimum variability timescale of the emission observed by the GBM, we use a method that calculates the chi-square of a differentiated time series of a lightcurve by subtracting counts in an individual bin  from those of the adjacent bin \citep{1997JGR...102.9659N}. This technique does not assume any pulse shape and the estimated minimum variability timescale is 0.8$\pm$0.1 s and its lower limit is 0.56$\pm$0.04 s.  
Thus, we divided the time intervals into $\sim$ 1-s bins to consider the temporal evolution and fitted the obtained spectra with Band+PL and DISKPBB+PL.  The best-fitting parameters for each model are summarized in Table \ref{spec_result_fine_resolve}. The series of the energy spectra fitted with  DISKPBB+PL for the divided  intervals  are shown in Figure \ref{fig:spec_time_resolved_pow_BB}.
We compared the results obtained  with Band+PL and  DISKPBB+PL as shown in Table \ref{spec_result_fine_resolve} and 
 found that both models give acceptable fitting results  ($\Delta$PGstat $\leq$ 4).

\begin{figure}[h]
\epsscale{1.00}
\plotone{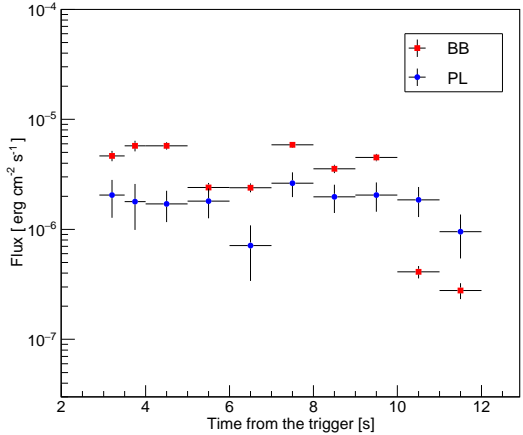}
\caption{Temporal evolution of the energy flux of the  BB and  PL components during intervals {\it b} and {\it c} where the additional power-law component is significantly detected. The displayed errors correspond to 1 $\sigma$ confidence.\label{flux_BB_nth}}
\end{figure}

Figure \ref{flux_BB_nth} shows the temporal evolution of the energy flux of the  BB and PL components. 
While the energy flux of the  PL component  seems to be almost constant during the displayed interval, the  BB component shows large variability and there is  likely to be no correlation between the  BB and PL components. This  may indicate that  the origin (i.e., emission site or/and mechanism) of the  BB component is entirely different from that of the  PL component. We note that this is  not the same as the strong correlation between  BB and PL emission detected in GRB 090902B \citep{2010ApJ...709L.172R}.

\begin{table*}[h]
\begin{center}
\caption{Fine time-resolved spectral fitting in the time intervals {\it b} and {\it c}.\label{spec_result_fine_resolve}}
\begin{tabular}{clccccc}
\tableline\tableline
Interval & Model & $\alpha$, $p$ & $\beta$ & $E_{\rm peak}$, $kT_{\rm ob}$ & $\Gamma_{\rm ext}$ & PGstat/dof  \\
{[s]} & & &  &  [keV] &  &   \\
\tableline
2.9 -- 3.5 & Band+PL & +0.14$^{+0.52}_{-0.37}$ & $-$3.06$^{+0.61}_{-unc}$  & 937$^{+266}_{-187}$ & $-$1.87$^{+0.10}_{-0.14}$   & 508/480   \\
 &  DISKPBB+PL &  1.05$^{+1.13}_{-0.21}$ &   & 353$^{+114}_{-83}$ &  $-$1.85$^{+0.12}_{-0.19}$   &  509/481  \\   \hline
3.5 -- 4.0 & Band+PL & +0.40$^{+0.96}_{-0.39}$ & $-$2.80$^{+0.43}_{-unc}$  & 819$^{+275}_{-266}$ & $-$1.89$^{+0.11}_{-0.30}$   & 478/480   \\
 &  DISKPBB+PL &  1.16$^{+2.09}_{-0.27}$ &   &  334$^{+108}_{-82}$ &  $-$1.86$^{+0.23}_{-0.13}$   &  480/481  \\   \hline
4.0 -- 5.0 & Band+PL & +0.31$^{+0.32}_{-0.29}$ & $-$5.75$^{+2.56}_{-unc}$  & 985$^{+168}_{-117}$ & $-$1.86$^{+0.08}_{-0.09}$   & 531/480   \\
 &  DISKPBB+PL &  1.16$^{+0.73}_{-0.22}$ &   &  366$^{+75}_{-61}$ &  $-$1.86$^{+0.14}_{-0.10}$   & 531/481  \\   \hline
5.0 -- 6.0 & Band+PL & +0.30$^{+0.50}_{-0.43}$ & $-$7.18$^{fixed}$  & 668$^{+158}_{-108}$ & $-$1.83$^{+0.08}_{-0.09}$   & 517/481   \\
 &  DISKPBB+PL &  1.12$^{+2.32}_{-0.26}$ &   &  250$^{+76}_{-57}$ &  $-$1.82$^{+0.12}_{-0.09}$   &  516/481  \\   \hline
6.0 -- 7.0 & Band+PL & +0.83$^{+0.66}_{-0.56}$ & $-$2.79$^{+0.43}_{-unc}$  & 566$^{+129}_{-97}$ & $-$1.95$^{+0.13}_{-0.40}$   & 503/480   \\
 &  DISKPBB+PL &  $>$ 1.07 &   &  209$^{+66}_{-42}$ &  $-$1.91$^{+0.27}_{-0.14}$   &  505/481  \\   \hline
7.0 -- 8.0 & Band+PL & +0.74$^{+0.51}_{-0.31}$ & $-$3.10$^{+0.53}_{-1.76}$  & 771$^{+97}_{-107}$ & $-$1.76$^{+0.05}_{-0.07}$   & 475/480   \\
 &  DISKPBB+PL &  $>$ 1.29 &   &  271$^{+52}_{-31}$ &  $-$1.76$^{+0.08}_{-0.06}$   &  478/481  \\   \hline
8.0 -- 9.0 & Band+PL & +0.72$^{+0.67}_{-0.61}$ & $-$2.61$^{+0.31}_{-2.60}$  & 635$^{+166}_{-110}$ & $-$1.93$^{+0.10}_{-0.31}$   & 536/480   \\
 &  DISKPBB+PL &  $>$ 1.01 &   &  261$^{+67}_{-48}$ &  $-$1.89$^{+0.12}_{-0.09}$   &  536/481  \\   \hline
9.0 -- 10.0 & Band+PL & +0.67$^{+0.65}_{-0.47}$ & $-$2.54$^{+0.23}_{-1.29}$  & 685$^{+138}_{-130}$ & $-$1.94$^{+0.12}_{-0.24}$   & 524/480   \\
 &  DISKPBB+PL &  $>$ 1.06 &   &  271$^{+65}_{-26}$ &  $-$1.84$^{+0.10}_{-0.08}$   &  529/481  \\   \hline
10.0 -- 11.0 & Band+PL & +0.94$^{+2.09}_{-0.79}$ & $-$6.74$^{fixed}$  & 255$^{+60}_{-49}$ & $-$1.84$^{+0.06}_{-0.08}$   & 476/481   \\
 &  DISKPBB+PL &  $>$ 0.91 &   &  84$^{+43}_{-14}$ &  $-$1.84$^{+0.09}_{-0.07}$   &  475/481  \\   \hline
11.0 -- 12.0 & Band+PL & +2.07$^{+1.94}_{-1.97}$ & $-$6.52$^{fixed}$  & 211$\pm$13 & $-$1.85$^{+0.09}_{-0.10}$   & 486/481   \\
 &  DISKPBB+PL &  $>$ 0.70 &   &  76$^{+64}_{-16}$ &  $-$1.84$^{+0.15}_{-0.10}$   &  486/481  \\   \hline

\tableline
\end{tabular}
\tablecomments{Notations are presented in the same manner in Table \ref{spec_result_exp_comp}. }
\end{center}
\end{table*}

\section{Discussion}\label{Sec:Discussion} 
\subsection{Constraint on the bulk Lorentz factor $\eta$}\label{Sec:ConstraintGamma}
High-energy gamma-ray photons could be absorbed by lower-energy photons through a process of pair creation and the opacity depends on the bulk Lorentz factor $\eta$ of the relativistic shell.
While the thermal photons may be emitted from an inner radius,
high-energy photons belonging to the power-law component may be emitted
from an outer radius of $R \simeq c \Delta t \eta^2/(1+z)$.
In the time interval ``$b$", the dominant target photons for $\gamma
\gamma$-collision
are the thermal component, for whose spectrum we adopt an approximate function
$f(E,T)\simeq6\times10^{-8}(E/\mbox{keV})^2/(\exp{(E/kT)}-1)$ photons/$\mbox{cm}^2$/keV/s, where $kT=250$ keV.
Then, the photon density in the jet comoving frame is estimated as
\begin{equation}
n'_{\rm th}(E')=\left( \frac{d_{\rm L}(z)}{R} \right)^2
\frac{\eta f(E,T) \Delta t}{(1+z)^3 W'},
\end{equation}
where $E'=(1+z)E/\eta$, $d_{\rm L}(z)$ is the luminosity distance,
and $W'$ is the width of the emission region in the comoving frame.
With this density, the optical depth $\tau(E'_0)\simeq \int dE' n'_{\rm th}(E')
\sigma_{\gamma \gamma}(y)W'$, where the cross section $\sigma_{\gamma \gamma}$
is a function of $y=\sqrt{1-2(m_{\rm e}c^2)^2/(E'_0 E')}$ \citep{2009Sci...323.1688A},
provides the minimum Lorentz factor $\eta_{\rm min}$ corresponding
to $\tau(E'_0)=1$
for $E_0=\eta E'_0/(1+z)=3.4$ GeV.

Here, we apply the same procedure described in Sec. \ref{Sec:FineTimeResolvedSpec} to the LAT data above 20 MeV and we find that the variability timescale is  $\Delta$t $>$ 0.6 s. If we  assume that {\it z} = 2,  which is the average value for long GRBs,  the minimum bulk Lorentz factor required in order to detect a  photon with an energy of 3.4 GeV, is   $\eta_{\rm min}$ $=$ 340.

In addition, if we assume a conservative timescale of $\Delta$t = 1.2 s, which is derived by the shortest-duration significant feature in the LAT light curve,  we obtain $\eta_{\rm min}$ $=$ 200.

\subsection{Properties of the photosphere}\label{Sec:Photosphere}
The obtained values for the temperature and flux of the thermal component in Section \ref{Sec:FineTimeResolvedSpec} can give 
us  useful information about the physical parameters of the photosphere.
\cite{2007ApJ...664L...1P} introduced a dimensionless parameter $\mathcal{R}$ using the observed blackbody flux $F_{\rm BB}^{\rm ob}$ and temperature  $kT_{\rm ob}$  for a GRB at redshift {\it z} and corresponding luminosity distance $d_L$, to characterize the bulk Lorentz factor $\eta$, photospheric radius $r_{\rm ph}$ and initial radius of the fireball $r_{\rm 0}$ at which the relativistic expansion begins. 

\begin{equation}
\mathcal{R} =  \left(\frac{F_{\rm BB}^{\rm ob}}{\sigma_{\rm ST} T_{\rm ob}^4}\right)^{1/2} = \frac{(1+z)^2}{d_L}\frac{r_{\rm ph}}{\eta} 
\label{eq:dimensionlessR}
\end{equation}
where $\sigma_{\rm ST}$ is  the Stefan-Boltzmann constant and we omit a geometrical factor of order unity which does not affect our conclusion significantly.
 While \cite{2007ApJ...664L...1P} considered a pure fireball scenario without any magnetization, \cite{2013A&A...551A.124H} improved and generalized the case of the magnetized outflow. The revised bulk Lorentz factor $\eta$, photospheric radius $r_{\rm ph}$, and initial radius $r_{\rm 0}$ are written as 

\begin{equation}
\eta = \biggl[  \left(1+z\right)^2 \left(1-\phi\right) \frac{\sigma_{\rm T} d_L F^{\rm ob}_\gamma }{2 m_{\rm p}c^3\mathcal{R}} \biggr]^{1/4} \biggl[ \frac{\epsilon_{\rm TH}}{f_{\rm NT}} \biggr]^{1/4} ,
\label{eq:Gamma}
\end{equation}

\begin{equation}
r_{\rm ph} = \biggl[  \left(1-\phi\right) \frac{\sigma_{\rm T} d_L ^5 \mathcal{R}^3 F^{\rm ob}_\gamma}{  2 m_{\rm p}c^3 \left(1+z\right)^6} \biggr]^{1/4} \biggl[ \frac{\epsilon_{\rm TH}}{f_{\rm NT}} \biggr]^{1/4} 
\label{eq:rph}
\end{equation}
and
\begin{equation}
r_{\rm 0} = \mathcal{R}\frac{d_{\rm L}}{(1+z)^2}\Bigl[ \frac{F^{\rm ob}_{\rm BB}}{ \left(1-\phi\right) F^{\rm ob}_\gamma} \Bigr]^{3/2} \biggl[ \frac{f_{\rm NT}} {\epsilon_{\rm TH}}\biggr]^{3/2} 
\label{eq:r0}
\end{equation}
where $F^{\rm ob}_{\rm \gamma}$ is the observed gamma-ray flux (e.g.,   DISKPBB + PL, $F^{\rm ob}_{\rm \gamma}$  = $F_{\rm BB}^{\rm ob}$ + $F_{\rm PL}^{\rm ob}$),  $m_{\rm p}$ is the proton mass, $\sigma_{\rm T}$ is the Thomson cross section, $c$ is the speed of light, $\phi$  
$\equiv$ $F_{\rm BB}^{\rm ob}$/$F^{\rm ob}_{\rm \gamma}$,  $f_{\rm NT}$ is the efficiency of the unknown process contributing to non-thermal emission from the injected total power. The ratio between the total injected power to the thermal radiation is $\epsilon_{\rm TH}$. The remaining power is converted to magnetic power.
As shown in Equations \ref{eq:Gamma}, \ref{eq:rph} and \ref{eq:r0}, $\eta$, $r_{\rm ph}$ and $r_{\rm 0 }$ include unknown parameters of $f_{\rm NT}$ and $\epsilon_{\rm TH}$.  The case of  $f_{\rm NT}$ = 1 is an extreme situation and quite unrealistic, so that $f_{\rm NT}$ $\leq$ 0.1 is favored \citep{1998MNRAS.296..275D}. 
The formalization assumes that the acceleration of outflows is complete  and there is no significant energy dissipation below the photospheric radius.

\begin{figure*}[h]
\epsscale{2.00}
\plotone{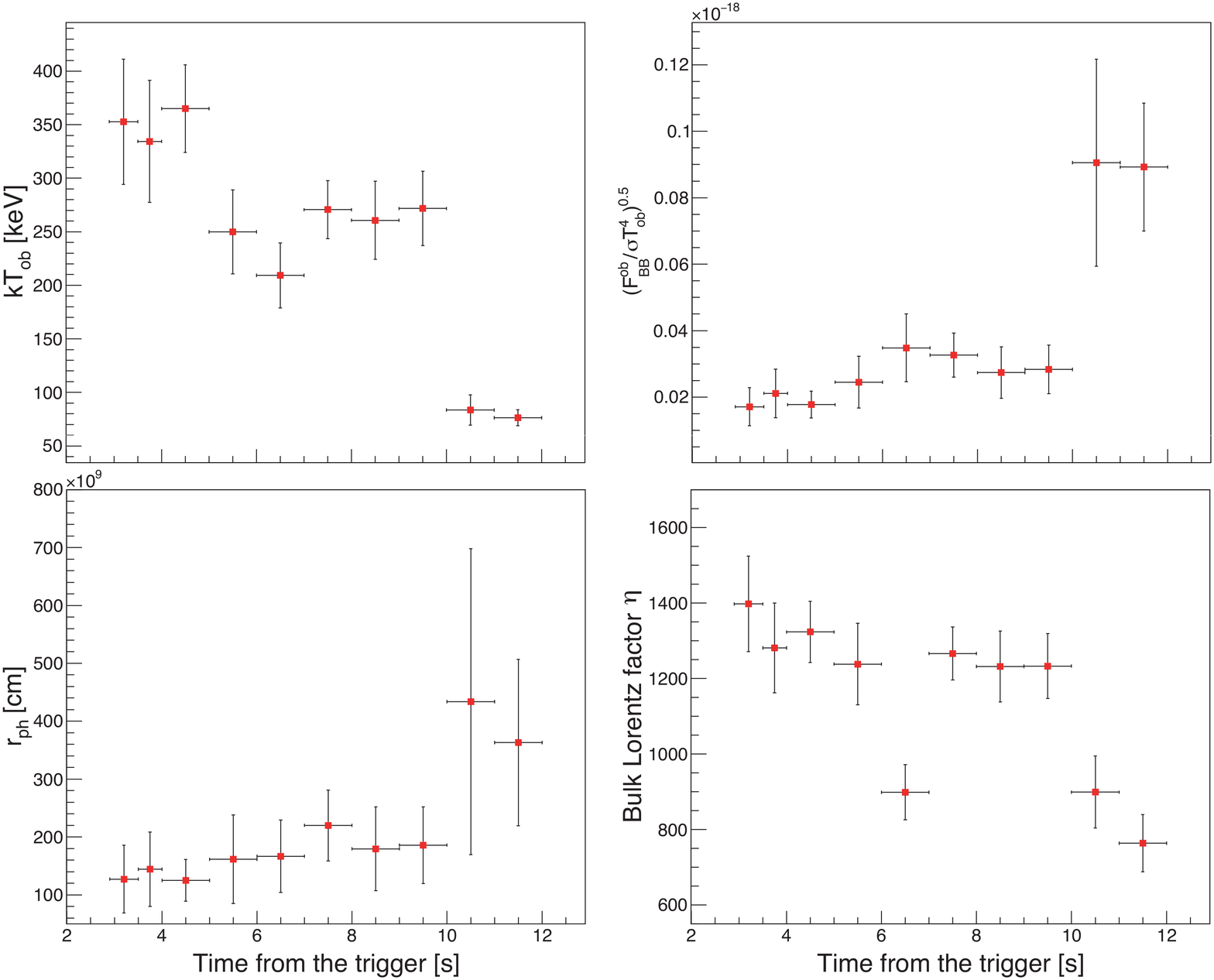}
\caption{Blackbody temperature $kT_{\rm ob}$, dimensionless size $\mathcal{R}$ =  ($F^{\rm ob}_{\rm BB} / \sigma_{\rm ST} T_{\rm ob}^4)^{1/2}$, 
photospheric radius $r_{\rm ph}$ and bulk Lorentz factor $\eta$  as a function of time during intervals {\it b} and {\it c} where the additional power-law component is significantly detected, assuming that the redshift of the burst is 2 (average value of long GRBs).  When calculating $r_{\rm ph}$ and $\eta$, we set $f_{\rm NT}$ = 0.1  and $\epsilon_{\rm TH}$ = 1. The details are described in the text.   
 The displayed errors correspond to 1 $\sigma$ confidence.}\label{fig:time_R_rph_gamma}
\end{figure*}

Figure \ref{fig:time_R_rph_gamma} shows the temporal evolution of the blackbody temperature $kT_{\rm ob}$, $\mathcal{R}$, $r_{\rm ph}$ and $\eta$, 
 assuming that  $f_{\rm NT}$ = 0.1, $\epsilon_{\rm TH}$ = 1 and $z$ = 2. The esimates of $\eta$ and $r_{\rm ph}$ do not strongly depend on the assumption of $f_{\rm NT}$ and $\epsilon_{\rm TH}$, and the obtained values do not change greater than an order of magnitude.
The obtained value of the temperature $kT_{\rm ob}$ and the bulk Lorentz factor $\eta$ are almost constant during  the initial 10 s from the trigger time although some dispersion is seen, after which the temperature and the bulk Lorentz factor decrease. The value of $\mathcal{R}$ is also constant during the early period and then rapidly increases. The photospheric radius is almost constant during the whole analyzed interval.

 Some GRBs  that have a thermal component show a temporal evolution of $T_{\rm ob}$ that is well described by a broken power law. In our case, we do not find the same behavior.
As shown  in Equation \ref{eq:dimensionlessR}, the physical parameter $\mathcal{R}$ is interpreted as an effective crossing size of the emitting region of the photosphere taking into account relativistic effects. Thus, a constant $\mathcal{R}$ implies a situation where the effective crossing size does not depend on time   and  $F_{\rm BB}^{\rm ob}$ $\propto$  $T^{4}_{\rm ob}$, which is the fundamental relation for a blackbody emitter.  During the initial 10-s interval, the estimated $F_{\rm BB}^{\rm ob}$ and $T_{\rm ob}$ for GRB 141207A seem to follow this relation. In addition,  some GRBs exhibit a monotonically increasing value of $\mathcal{R}$ with time \citep{2009ApJ...702.1211R}, which is similar to the increase of  $\mathcal{R}$  towards the late time emission of GRB141207A.

  The initial radius $r_{\rm 0}$ is almost constant during the whole analyzed interval with an average value of $r_{\rm 0}$ $\sim$ 4 $\times$ 10$^6$ ($f_{\rm NT}$/0.1)$^{3/2}$ ($\epsilon_{\rm TH}$/1)$^{-3/2}$ cm ($r_{\rm 0}$ = 10 -- 400  ($\epsilon_{\rm TH}$/1)$^{-3/2}$ km for  $f_{\rm NT}$ = 0.05 -- 0.5). 
The radius $r_{\rm 0}$ is an important parameter to deduce a progenitor of this burst.  We first compare the derived value to the variability timescale of this burst. The obtained timescale gives us the light crossing length $r^\prime$ which might be an indicator of $r_{\rm 0}$ and the variability timescale  $\delta t$ $>$  $\sim$ 0.6 s as described in Section \ref{Sec:ConstraintGamma}, giving $r^\prime$ $>$ $c\delta t$ $\sim$ 10$^{10}$ cm. This is far larger than the obtained value of $r_{\rm 0}$  even if we choose the extreme case of $f_{\rm NT}$ =1. Although an extreme case of $\epsilon_{\rm TH}$ $\ll$ 1 (i.e., magnetically dominated flow) can also obtain $r_{\rm 0}$ $\sim$ 10$^{10}$ cm,  such a small $\epsilon_{\rm TH}$ does not agree with the observed thermal  to non-thermal flux ratio  \citep{2013A&A...551A.124H}. The obtained timescale may be controlled by the variability of the accretion onto the central engine rather than the light crossing time.

Next, we consider a case  where the initial radius $r_{\rm 0}$ is possibly associated with the gravitational binding radius $r_{\rm g}$, 
 which is sometimes assumed to be the innermost stable orbit around a black hole at $r_{\rm g}$ = 3  $r_{\rm sch}$, where $r_{\rm sch}$ is the Schwarzschild radius. 
 Assuming a massive progenitor of 30 ${\rm M}_\sun$ \citep{1995ApJS..101..181W} the expected gravitational radius $r_{\rm g}$ would then be $\sim$ 3 $\times$10$^7$ cm. 
  Although $r_{\rm 0}$ could  be equal to $r_{\rm g}$ when adopting $f_{\rm NT}$ $\sim$ 0.5  and $\epsilon_{\rm TH}$ = 1, the used parameters have  large uncertainties especially for  $f_{\rm NT}$.
 We note that when choosing very small $f_{\rm NT}$ $<$ 0.01 we find the injection radius to be $r_{\rm 0}$ $<$ 1 km, which is challenging,  even if we adopt $\epsilon_{\rm TH}$ = 0.1 -- 1.

\subsection{Afterglow}\label{sec:afterglow}
With the  datasets of the LAT temporally extended emission and the X-ray afterglow detected by the XRT, we tested whether the external forward shock (ES) model \citep{1998ApJ...497L..17S} can explain the temporal and spectral behaviors at the late time. Unfortunately there are not many simultaneous observations with LAT and XRT.

\begin{figure}[h]
\epsscale{1.0}
\plotone{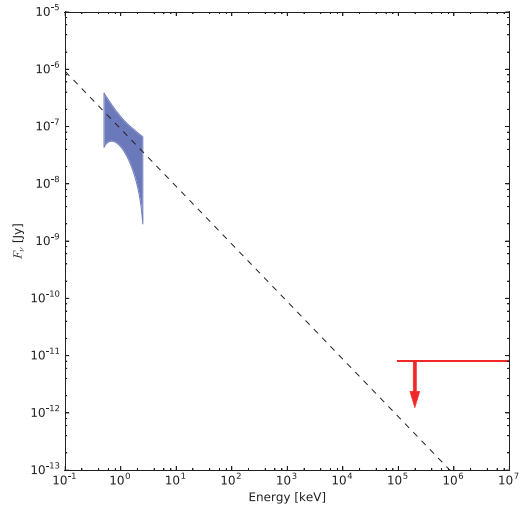}
\caption{SED of the afterglow of GRB 141207A with XRT (the blue filled area at the 1-$\sigma$ confidence level) and LAT data (the red arrow at the 95\% confidence level) from $T_0$ + 46 ks to 63 ks. The dashed line is the best-fit power-law function with the photon index of $\Gamma$ = $-$2.00$\pm$0.46. \label{SED_afterglow}}
\end{figure}

 The flux evolution in the afterglow phase is described using $F_\nu$ $\propto$ $t^{\alpha_{\rm AG}}$ $\nu^{\beta_{\rm AG}}$, where the photon index $\Gamma$ = $\beta_{\rm AG}$ $-$1. 
The obtained temporal decay indices ($\alpha_{\rm AG}$) for the LAT and XRT are  $\alpha_{\rm AG}^{\rm LAT}$ =  $-$1.5$\pm$0.2 and
 $\alpha_{\rm AG}^{\rm XRT}$ =  $-$1.9$\pm$0.6, respectively. 
  If we assume an interstellar medium (ISM) afterglow scenario in  slow-cooling phase, where the observed frequency $\nu_{\rm obs}$ is larger than the cooling frequency $\nu_{\rm c}$ (i.e., $\nu_{\rm c}$ $<$ $\nu_{\rm obs}$),
the expected photon indices are given by $\Gamma_{\rm rad}$ = (7$\alpha_{\rm AG}$ $-$ 14)/12 and $\Gamma_{\rm ad}$ = (2$\alpha_{\rm AG}$ $-$ 4)/3 for the radiative and adiabatic cases, respectively. Inserting the obtained LAT temporal index $\alpha^{\rm LAT}_{\rm AG}$, the expected photon indices are $\Gamma^{\rm LAT}_{\rm rad}$ = $-$2.0$\pm$0.1 and $\Gamma^{\rm LAT}_{\rm ad}$ = $-$2.3$\pm$0.1. The observed photon index is in the range of $-$2.2 to $-$1.2 as shown in Figure \ref{EE_LAT_Swift}, so the adiabatic index lies slightly outside the 1-$\sigma$ confidence level. The radiative jet might thus be slightly favored, although it is not significant. 
From the observed result obtained by XRT ($\alpha^{\rm XRT}_{\rm AG}$ =  $-$1.9$\pm$0.6), the expected photon indices for the radiative and adiabatic cases are $\Gamma^{\rm XRT}_{\rm rad}$ = $-$2.3$\pm$0.7 and  $\Gamma^{\rm XRT}_{\rm ad}$  = $-$2.6$\pm$0.8. The time-integrated spectrum  measured with XRT has a photon index of $-$2.0$_{-0.8}^{+0.7}$, which is consistent with both the radiative and adiabatic cases due to its large uncertainty.
 These observed and expected photon indices are summarized in Table \ref{AG_param}.
Furthermore,  we made the spectral energy density (SED) using the  simultaneous observational data with LAT and XRT at the defined time interval ($T_0$  +46 ks -- 63 ks) as shown in Figure \ref{EE_LAT_Swift}. The obtained SED  in Figure  \ref{SED_afterglow}  is well fitted  by a single power-law function with best-fit index of $\Gamma$ =  $-$2.0$\pm$0.5, which indicates that the canonical afterglow model is compatible with the observational results of GRB 141207A in the wide energy band.

 Furthermore, we test whether the physical parameters for ES model are valid. First we define $\epsilon_e$ and $\epsilon_B$ as the fractions of electron and magnetic energy transferred from the shock energy $E_{\rm ES}$, respectively, and  $n_{\rm ISM}$ as the density of the interstellar medium. Then
for the radiative cooling jet model \citep{1998ApJ...497L..17S}, we set  $\epsilon_e$ = 0.8, $\epsilon_B$ = 0.1,   $n_{\rm ISM}$ = 1 cm$^{-3}$, $\alpha_{\rm AG}$ = -1.6, and we choose $\eta$ = 1000 from our estimation in Sec. \ref{Sec:Photosphere}.
In order to obtain the observed LAT flux ($\sim$8 $\times$ 10$^{-7}$ photons/cm$^2$/s) at 1000 s after the GBM trigger (Figure \ref{EE_LAT_Swift}),
  we estimate that $E_{\rm ES}$ has to be 5 $\times$ 10$^{53}$ erg, assuming {\it z} = 2. Considering the observed non-thermal gamma-ray isotropic energy $E_{\gamma, {\rm iso}}^{\rm NT}$ = 2.3 $\times$ 10$^{53}$ erg, the efficiency responsible for the non-thermal emission is  $E_{\gamma, {\rm iso}}^{\rm NT}$/$E_{\rm ES}$ $\sim$ 0.5, which does not contradict the discussion of the photospheric emission in Sec \ref{Sec:Photosphere}.
 Under those conditions, we calculated the XRT flux at 5 $\times$ 10$^{4}$ s after the GBM trigger to be 5 $\times$ 10$^{-4}$ photons/cm$^2$/s, which is consistent with the observed XRT flux (4.8$\pm$1.2 $\times$ 10$^{-4}$ photons/cm$^2$/s). When we consider the case of the adiabatic model, the estimated XRT flux is too high ($\sim$ 1 $\times$ 10$^{-2}$ photons/cm$^2$/s), which seems to be incompatible with the observed flux. Thus, we find that our assumption for $\epsilon_e$, $\epsilon_B$ and   $n_{\rm ISM}$ of the radiative case is feasible and the obtained results strengthen the validity of the radiative afterglow model.

 The  maximum synchrotron photon energy in the radiative afterglow model  is derived  with the same procedure as in \cite{2014Sci...343...42A}. The highest-energy photon with an energy of 5.5 GeV was observed at $T_0$ + 734 s.
The observable highest synchrotron energy at $T_0$ + 734 s is calculated to be 3.8 GeV which is slightly smaller than that of the observed photon. However, a more stringent limit could be imposed by multiplying the calculated energy by a factor of 2$\pi$. In such a case, the observed highest-energy photon does not contradict the afterglow model, although the observed highest-energy photon requires extreme conditions for particle acceleration \citep[e.g.,][]{2014ApJ...788...36B}.
\begin{table}[t]
\begin{center}
\caption{Comparisons of Observed and Expected Photon indices $\Gamma_{\rm AG}$ for XRT and LAT.\label{AG_param}}
\begin{tabular}{cc|cc}
\tableline\tableline
\multicolumn{2}{c|}{Observed} & \multicolumn{2}{|c}{Expected}  \\
 & & &    \\
\tableline
$\Gamma^{\rm LAT}_{\rm AG}$ & $-$2.2 -- $-$1.2\tablenotemark{a} & $\Gamma^{\rm LAT}_{\rm ad}$ & $-$2.3$\pm$0.1   \\
& & $\Gamma^{\rm LAT}_{\rm rad}$ & $-$2.0$\pm$0.1   \\ \hline
$\Gamma^{\rm XRT}_{\rm AG}$ & $-$2.0$^{+0.7}_{-0.8}$ & $\Gamma^{\rm XRT}_{\rm ad}$ & $-$2.6$\pm$0.8   \\
& & $\Gamma^{\rm XRT}_{\rm rad}$ & $-$2.3$\pm$0.7   \\

\tableline
\end{tabular}
\tablenotetext{a}{ Time dependencies of the LAT photon indices are shown \\
 in Figure \ref{EE_LAT_Swift}}
\tablecomments{
 The expected photon indices for the adiabatic- \\
 and radiative-jet cases are calculated using the observed \\
 temporal decay indices $\alpha^{\rm LAT}_{\rm AG}$ with LAT and $\alpha^{\rm XRT}_{\rm AG}$ with XRT.}
\end{center}
\end{table}

\subsection{Origin of the power-law component}
\begin{figure}[t]
\epsscale{1.0}
\plotone{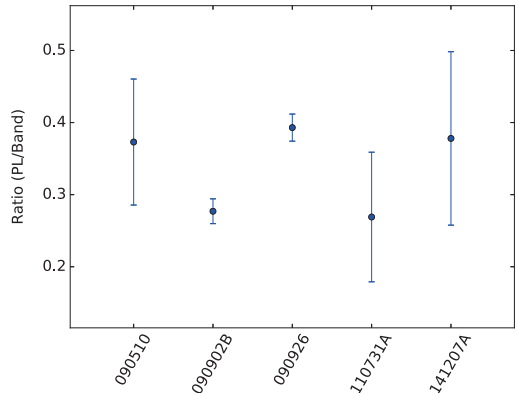}
\caption{Ratio of the additional power-law flux to the Band flux in the time-integrated spectrum of several GRBs including GRB 141207A. \label{HadronRatio}}
\end{figure}

The {\it Fermi} observation reveals the existence of an extra PL component to explain the excess emission at both low ($\sim$ 10 keV) and high  energies (GeV) in the prompt emission,  as shown in the residuals of the Band function alone of Figure \ref{specSummaryBand}. We estimate the flux contribution of the extra PL component relative to the Band component, and show this ratio together with that of  bright GRBs having an extra PL component (with or without exponential cutoff) in Figure \ref{HadronRatio}. The calculations were performed using the time-integrated spectra. The ratio is 30 $\sim$ 40 \% in all the GRBs including GRB 141207A. In addition, the photon index of the extra PL component $\Gamma_{\rm ext}$ is $\sim$ $-$1.9, which is also a similar to that seen in the
other GRBs except for the short GRB 090510 ($\Gamma_{\rm ext}$ $\sim$ $-$1.6). These features might indicate that the extra PL component originates from the same emission mechanism. 

The physical origin of the extra PL component is still under debate, but  both the internal and external shock origins are likely. For the internal-shock origin, there are two possibilities: leptonic and hadronic models. 

The leptonic models produce the GeV emission via synchrotron self-Compton emission and/or Comptonization of thermal photons by electrons, e.g., \cite{2010ApJ...720.1008C}, \cite{2011ApJ...739..103A}, \cite{2012MNRAS.420..468P}.   
 In the case of GRB 141207A, the GBM emission originates from the photosphere and the photosphere region is much smaller than the internal shock region which would contribute to the LAT emission. Although these regions are not co-spatial, the thermal photon from the photosphere could play the role of the seed photon responsible for the LAT emission via inverse Compton scattering. Under this assumption,
 we would expect that the high-energy emission in the (sub-) GeV band should be temporally correlated with the seed photon emission around the sub-MeV peak.
 As shown in  Figure \ref{flux_BB_nth}, the thermal component  seems to be highly variable with time while the PL component is  almost independent of time.  In order to estimate the correlation between the thermal and non-thermal components, we calculate the correlation coefficient, obtaining a value of 0.37$\pm$0.22, with a probability of 16\% that there is no correlation. This result is not very constraining because of large uncertainties and limited photon statistics especially for the LAT. Therefore, we can not claim that there exists no correlation between the thermal and non-thermal components and the leptonic scenario cannot be rejected. However, the low-energy excess below the peak energy of the thermal component is difficult to explain via an inverse Compton scenario.

For the hadronic models, accelerated high-energy protons originating from the internal shocks trigger an electronic cascade caused by photopion production, e.g., \cite{2009ApJ...699..953A, 2009ApJ...705L.191A, 2010ApJ...725L.121A}. The cascade emission could easily  realize an additional PL component with a flat spectrum (photon index $\sim$ $-$1.9) in a very broad energy band, from a few keV to a few GeV. Furthermore, a strong magnetic field can be invoked for high photo-meson production efficiency in order to avoid an extremely large proton luminosity, while for the leptonic models a large magnetic field drastically decreases the contribution of the inverse Compton component.
In addition, as the proton acceleration timescale is much longer than the electron cooling time, 
 the variability timescale of the PL component may be broadened \citep{2012ApJ...757..115A}. 
Although the photon statistics are not significant to verify the
broadness of the pulse profile in the GeV light curve,
the hadronic model represents a valuable choice for interpreting  this GRB.

Besides the internal shock scenario, we suggest the ES scenario, e.g., \cite{2010MNRAS.409..226K}: synchrotron emission is produced via an external forward shock and the LAT emission originates from the early onset of the afterglow.
As described in Section \ref{sec:afterglow}, the obtained spectral and temporal parameters are consistent with the ES model and the fact that the flux variability of the high-energy PL component does not  seem to depend on time also may favor the ES model.
Furthermore, the ES model could explain the delay of the (sub-)GeV emission with respect to the X-ray GBM emission, corresponding to the deceleration onset of the ES ($t_{\rm onset}$). From \cite{1999ApJ...520..641S} we obtain the deceleration radius as follows, 

\begin{equation}
t_{\rm onset} = \biggl[   \frac{3E_{\rm ES} (1+z)^3}{32 \pi n_{\rm ISM}m_{\rm p}c^5  \eta^8 } \biggr]^{1/3}
\label{eq:t_onset}
\end{equation}

If we take $z$ = 2
and $n_{\rm ISM}$ $=$ 1 cm$^{-3}$, we obtain

\begin{equation}
t_{\rm onset, z=2} = 2 \;{\rm s} \left(  \frac{E_{\rm ES}}{ 5 \times 10^{53} \;{\rm erg}} \right)^{1/3} \left(  \frac{ n_{\rm ISM}}{ 1 \;{\rm cm}^{-3}} \right)^{-1/3}  \left(  \frac{ \eta}{1000} \right)^{-8/3}  
\label{eq:t_onset_z2}
\end{equation}
The expected onset times are almost consistent with  the observed temporal delay ($\sim$ 3 s), which might  support a scenario where the LAT emission in both the prompt and temporally extended phases  is explained by the ES model.   The observed variability timescale in Sec \ref{Sec:ConstraintGamma} is typically $\sim$1 s and this is almost consistent with the deceleration timescale within an order of magnitude. 
However, we note that as shown in Figure \ref{EE_LAT_Swift} the observed photon index at late phase ($T_0$ + 1000 s) is larger than the expected one, although its uncertainty is large.   This inconsistent signature with the standard afterglow scenario may indicate additional components other than the external forward shock.

\subsection{Very high-z burst ?}\label{sec:highzburst}
 We have assumed that {\it z} = 2 so far and we try to estimate the redshift for this GRB using X-ray and gamma-ray data.
To estimate the  distances of GRBs with unknown redshift, the empirical relationships between the peak energy $E_{\rm peak}$ and the isotropic equivalent luminosity $L_{\rm iso}$  could be useful, although 
the physical origin of this relationship is uncertain and conclusions drawn from it need to be treated with great caution. For example, some  studies claim that such relations might be due to observational selection bias \citep[e.g.,][]{2012MNRAS.422.2553G}. While the Yonetoku relation \citep{2004ApJ...609..935Y, 2010PASJ...62.1495Y} has been widely used when fitting the GRB spectrum with a single Band function, for some GRBs with additional PL component it does not give a suitable result due to inclusion of the extra  PL component.  \cite{2015ApJ...807..148G} suggested  treating  the Band and PL components separately and use only the contribution of the Band component when establishing the empirical relationship.

\begin{figure}[t]
\epsscale{1.0}
\plotone{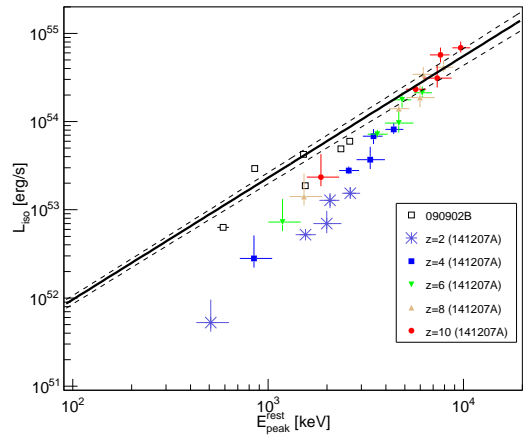}
\caption{  The plot of the peak energy $E_{\rm peak}^{\rm rest}$ of the Band function (or of the cutoff power-law function) at the rest frame of GRB 141207A  versus isotropic luminosity $L_{\rm iso}$ assuming $z$ = 2, 4, 6, 8, 10, when fitting the Band function (or the cutoff power-law function)  + the power-law function at the coarse time intervals as presented in Section \ref{SubSec:CoarseTimeResolvSpec}.  The white squares represent GRB090902B with a fit of the Band function + the  power-law function. The solid black line and dashed black lines represent  the best-fit power-law function  and the $\pm$1$\sigma$ ranges derived from  \cite{2015ApJ...807..148G}, respectively.
 \label{Fig:Guiriec}}
\end{figure}

 We decide to adopt the approach of \cite{2015ApJ...807..148G} to estimate the possible redshift of GRB 141207A and we assume that the thermal component, which is also well fitted by the Band function,  is adopted as a spectral component used for the empirical relation. This assumption is quite strong;  \cite{2015ApJ...807..148G} fitted observed spectra with Band + BB + PL, where the BB component was  sub-dominant. They then used the dominant Band component (representing the non-thermal emission) for their empirical relation. In our case, we assume that the Band component comes from photospheric emission during which the additional PL component exists. To test whether this assumption is valid, we check the validity of the empirical relation to GRB 090902B with known redshift 
 that has a prominent photospheric component well represented by the Band function plus a non-thermal PL component or a modified BB plus a PL component; the spectral feature seen in GRB090902B is very similar to the one observed in GRB141207A. The obtained result is shown in Figure \ref{Fig:Guiriec}. We see that the Band component for GRB 090902B seems to be consistent with the empirical relation by  \cite{2015ApJ...807..148G} and find that our assumption is not unreasonable. 
 
 The $E_{\rm peak}$ -- $L_{\rm iso}$ plot for the coarse time-resolved intervals in Section \ref{SubSec:CoarseTimeResolvSpec} assuming several redshifts ($z$ = 2 -- 10)  is shown in Figure \ref{Fig:Guiriec}.
When deriving the spectral parameter of the Band component we adopted the Band function (or the power-law function with exponential cutoff)  plus the power-law function. Furthermore,  since we do not see any statistical improvement by adding an extra component in the initial time interval  from $T_0$ $-$0.5 s to +2.9 s, we adopt the single Band function. 
 We find that 
the most probable redshift  is $z$ =  10.3 and $z$ $>$  6.5 with 3-$\sigma$  confidence level, indicating that GRB 141207A might have occurred in the very high-$z$ universe. Amazingly the obtained redshift ($z$ $\sim$ 10) nearly corresponds to the reionization epoch \citep{2011ApJS..192...18K}, and could directly indicate that GRB 141207A belongs to the first sources of light (i.e., Pop III stars).   We note that, although the slope of our data points is slightly steeper than that by \cite{2015ApJ...807..148G} as shown in Figure \ref{Fig:Guiriec}, this might be due to a scatter around the relation. Indeed,  some scatter is also seen in  \cite{2015ApJ...807..148G}. In addition, even if we exclude the photospheric data points at the time intervals of ``$b$" and ``$c$", we still obtain a high-{\it z} value of {\it z} $>$ 8 with 3-$\sigma$  confidence level. 
 Recently, \cite{2016arXiv160607193G} found that the model of \cite{2015ApJ...807..148G} was also
successful in matching the spectroscopic redshift of GRB 110205A.
Additionally, they found that the spectral model matched also the optical data. However,
 in our case the optical data was not observed during the prompt phase
and there is no significant {\it u}-band detection by UVOT   (3000 $\AA$ -- 3900 $\AA$) in the afterglow phase.
At  $z$ $\sim$ 2, the UVOT band nearly corresponds to the wavelength of the Lyman series.
Therefore, for $z$ $>$ 2 the afterglow emission is expected to be very
weak and the UVOT observation is
consistent with the high-z scenario.

 Finally, we check if the very high-{\it z} GRB scenario ({\it z} = 10.3) is compatible with the arguments previously discussed.  First, the estimated minimum bulk Lorentz factor is $\eta_{\rm min}$ $=$ 1100 with the same procedure in Sec. \ref{Sec:ConstraintGamma} using the estimated high redshift.
The derived physical parameters from the photospheric emission (e.g., $\eta$, $r_{\rm ph}$ and $r_{0}$) change only within a factor of 4 and especially the obtained $\eta$ $\sim$ 3000 does not contradict  $\eta_{\rm min}$.
When we discuss the ES scenario, although we obtain the reasonable deceleration time of  $\sim$ 2 s ($E_{\rm ES}$/3.5 $\times$ 10$^{55}$ erg)$^{1/3}$ ($n_{\rm ISM}$/ 1 cm$^{-3}$)$^{-1/3}$ ($\eta$/3000)$^{-8/3}$,
 the required isotropic kinematic energy of $E_{\rm ES}$ = 3.5 $\times$ 10$^{55}$ erg seems to be extremely large and challenging.  Furthermore, the estimated maximum synchrotron energy under typical conditions (not multiplying by 2$\pi$) is 1.0 GeV at $T_0$ + 734 s and the observed photon with 5.5 GeV highly exceeds the limit of the radiative afterglow model.

In summary, while the very high-{\it z} scenario is partially consistent with results assuming that {\it z} = 2, 
there are some big challenges; e.g., for the ES model the required kinetic energy is extremely large ($\sim$ 10$^{55}$ erg) and the observed highest energy photon cannot be easily explained by the ES model. 
 We caution that the empirical relation of \cite{2015ApJ...807..148G} was derived using a
strong non-thermal component in the spectrum, whereas GRB 141207A appears to
be dominated by a thermal component. Such GRBs \citep[including for example GRB 090902B;][]{2010ApJ...709L.172R} 
 may belong to a different population of GRBs. In
addition, the relation of \cite{2015ApJ...807..148G} has not been tested for high-redshift
GRBs.

\section{Conclusion}\label{Sec:Conclusion}
We have presented the observation, detailed analysis results and physical interpretations for GRB 141207A. This burst has   features typically seen in other LAT-detected bursts, 
such as a delayed onset of the GeV emission with respect to the X-ray emission and temporally extended emission in the LAT range.
During the first two pulses in the prompt emission phase, the spectrum  of GRB 141207A is well fitted by 
a Band  component with hard low-energy photon index  or modified blackbody component plus an additional power-law component spanning a wide energy range from keV to GeV,  which indicates that this GRB has a photospheric origin. In addition,
our finely time-binned analysis shows that the estimated bulk Lorentz factor $\eta$ $\sim$ 1000 and the initial radius $r_0$ = 10 -- 400 km depends on the unknown non-thermal efficiency $f_{\rm NT}$.

 The physical origin of the non-thermal power-law component in the prompt phase is not understood, and we discussed  whether the component can be explained by the leptonic and hadronic scenarios of the internal shock regime and  synchrotron emission from external forward shock.  Due to the limited photon statistics, our temporal analysis is not able to distinguish those models, therefore we cannot find strong evidence for a favorable one. Observations of bright GRBs by {\it Fermi} will likely give us a key to disentangle these models.

The temporally extended LAT emission and the X-ray afterglow  were also detected and the results obtained by the temporal and spectral analyses support the radiative external forward shock scenario. 
However, in the late phase of the extended emission in the LAT band ($T_0$ + 1000 s) the spectra show signs of hardening with time, which cannot be explained by synchrotron emission in the standard scenario and another contribution may exist.

Finally, our redshift estimate using the empirical $E_{\rm peak}$ -- $L_{\rm iso}$ relationship of \cite{2015ApJ...807..148G}  may point to GRB 141207A occurring in the very high-z universe ($z$ $\sim$  10).
  For such a high-{\it z}  event,  an extremely large injection energy ($\sim$ 10$^{55}$ erg) is needed and there is an inconsistency with the external forward shock (e.g., the observed highest energy photon exceeds the maximum synchrotron energy). 
Note that we used a controversial empirical relation for estimating the redshift, which is not verified for such a special type of energetic (e.g., Pop III) GRBs at high-{\it z} universe.

\acknowledgments
 
The \textit{Fermi} LAT Collaboration acknowledges generous ongoing support
from a number of agencies and institutes that have supported both the
development and the operation of the LAT as well as scientific data analysis.
These include the National Aeronautics and Space Administration and the
Department of Energy in the United States, the Commissariat \`a l'Energie Atomique
and the Centre National de la Recherche Scientifique / Institut National de Physique
Nucl\'eaire et de Physique des Particules in France, the Agenzia Spaziale Italiana
and the Istituto Nazionale di Fisica Nucleare in Italy, the Ministry of Education,
Culture, Sports, Science and Technology (MEXT), High Energy Accelerator Research
Organization (KEK) and Japan Aerospace Exploration Agency (JAXA) in Japan, and
the K.~A.~Wallenberg Foundation, the Swedish Research Council and the
Swedish National Space Board in Sweden.

 Additional support for science analysis during the operations phase is gratefully acknowledged from the Istituto Nazionale di Astrofisica in Italy and the Centre National d'\'Etudes Spatiales in France.

This work is supported  by the Ministry of Education, Culture, Sports, Science and Technology (MEXT), Grant-in-Aid No. 24103002.







\appendix

\clearpage

\end{document}